\documentclass[12pt,preprint]{aastex}
\shorttitle{A New Definition for Giant Planets}
\shortauthors{Hatzes and Rauer}
\begin{document}

\title{A Definition for Giant Planets Based on the \\
Mass $-$ Density Relationship}

\author{Artie P. Hatzes}
\affil{Th\"uringen  Landessternwarte Tautenburg, Sternwarte 5, D-07778 Tautenburg, Germany}
\email{artie@tls-tautenburg.de}
\and
\author{Heike Rauer}
\affil{Institut f\"ur Planetenforschung, Deutsches Zentrum f\"ur
Luft- und Raumfahrt, Rutherfordstrasse 2, D-12489, Berlin, Germany}
\affil{Zentrum f\"ur Astronomie und Astrophysik, TU Berlin, Hardenbergstr. 36, D-10623 Berlin, Germany}
\email{Heike.Rauer@dlr.de}

\begin{abstract}
We present the mass-density relationship (log $M$ - log $\rho$) for objects
with masses ranging from planets ($M$  $\approx$ 0.01 $M_{Jup}$) through stars ($M$ $>$ 0.08 $M_\odot$).
This relationship shows three distinct regions separated by a change in 
slope in log $M$ -- log $\rho$ plane. In particular, objects with masses in the
range 0.3  $M_{Jup}$ to 60 $M_{Jup}$ follow a tight linear relationship 
with no distinguishing feature to separate the low mass end (giant planets) from the
high mass end (brown dwarfs). The distinction between giant planets and brown
dwarfs thus seems arbitrary. We propose a new definition of giant planets 
based simply on changes in the slope of the log $M$ versus log $\rho$ 
relationship. By this
criterion, objects with masses less than $\approx$ 0.3 $M_{Jup}$ are low mass
planets, either icy or rocky.
Giant planets cover
the mass range 0.3 $M_{Jup}$ to 60 $M_{Jup}$. Analogous to the stellar main
sequence,  objects on the upper end of the 
giant planet sequence (brown dwarfs) can simply be referred to as ``high mass giant planets", while
planets with masses near that of Jupiter can be considered to be 
``low mass giant planets''. 
\end{abstract}

\keywords{planets and satellites: fundamental parameters --- stars: low mass -- stars:  brown dwarfs}

\section{Introduction}
The nature of stellar and sub-stellar objects is determined by their mass.
Stars  are  defined
as an object with sufficient mass to  ignite
hydrogen fusion in the  core. 
Sub-stellar objects, on the other hand,
 have masses below that needed to ignite hydrogen burning
in their cores ($M$ $\sim$ 80 $M_{Jup}$).  Like
their stellar counterparts,  sub-stellar objects encompass
a wide range of masses 
from those that are accepted as 
planets, with masses of a few $M_{Jup}$,  to objects often considered to
be brown dwarfs with
masses of a few tens of $M_{Jup}$. 
The exact boundary in mass between what one considers
a ``planet'' and what one considers a ``brown dwarf'''' is blurred and is
still the subject of debate.

One definition of  giant planet is that it is a sub-stellar
 object that has not undergone deuterium burning
anytime during its life. By this criterion the boundary between
planets and brown dwarfs should be about 13 $M_{Jup}$ (Burrows et al. 2001).
However, this distinction seems arbitrary as the mass distribution for  companions below
25 $M_{Jup}$ show no characteristic features at this mass limit (Udry 2010).
Furthermore, the period of deuterium burning has little influence
on the future evolution of the brown dwarf. This is contrary to stars
where hydrogen burning under hydrostatic equilibrium significantly alters
the further evolution of the object. Chabrier
et al. (2014) argued that deuterium burning, or lack thereof, plays no role in either giant
planet or brown dwarf formation. They also pointed out that
these two types of objects ``might bear some imprints of their formation mechanism, notably in their mean density and
in the physical properties of their atmosphere.''

On the other hand, the intersection of the
mass distributions  of sub-stellar objects appear to have a 
distinctive dip around $M$ $\approx$ 25-30 $M_{Jup}$ (Udry 2010). Schneider et al. (2011)
attributed this dip as the boundary between the mass spectrum of planets, which
is decreasing with increasing mass, and the distribution of sub-stellar and low mass
stars which is increasing beyond this point.
The dip is also coincident with a possible break in the mass-radius relationship
for low mass and sub-stellar objects (Pont et al. 2005; Anderson et al. 2011)
which suggests a difference in the physical natures between objects
on either side of this boundary (Schneider et al. 2011). For these
reasons Schneider et al. (2011) {\it arbitrarily} (our emphasis)
assigned a maximum mass of 25 $M_{Jup}$ as the limit for including
objects in the {\it Exoplanet Encyclopaedia} (www.exoplanet.eu).
However, if an object lies near this 25 $M_{Jup}$ boundary we still do not know
its nature, i.e. to which distribution (planets or brown dwarfs) it actually 
belongs. 

The mean density versus mass relationship for planets
show a broad minimum around a mass of 0.3 $M_{Jup}$
(Rauer et al. 2014; Laughlin \& Lissauer 2015) which separates the
H/He dominated giant planets from low-mass planets of Neptune-mass or smaller.
The different slopes
in the density-mass plane of the two objects highlights the differences in structure
between the two classes of planets.
We extend the density-mass relationship through sub-stellar and stellar objects and
show that this also shows a change in slope marking the differences in structure between
giant planets/brown dwarfs and stellar objects. We argue for a definition of giant planets
based on the mass-density diagram.

\section{The Density versus Mass Relationship}

	We constructed a mass-density diagram for the full range of masses
covering planets through main sequence stars. Because all of these are
transiting/eclipsing systems these are relatively close pairs. In this sense
they can be treated as  a ``pseudo-homogenous'' sample.
For the giant planet data we largely restricted our sample to transiting planets
from the space based missions CoRoT and Kepler.
These space missions provide light curves with the best photometric
precision that produce the most accurate planet radii. This is particularly
important since the radius enters as the third power in the density.
Besides using space-based transit discoveries that provide the best
photometric precision, it is important to use transit light curve analyses
that were done in a consistent manner. Various investigators may use
different limb-darkening laws or apply different methods to filter out
intrinsic  stellar variability which may introduce more scatter in the results.
For this reason we took 
radius and mass values for the planets from the web-based TEPCat catalog
(http://www.astro.keele.ac.uk/jkt/tepcat/tepcat.html \\ and references therein)
as these were derived from a uniform analysis (see Southworth 2010,2011,2013).

There are few mass and radius measurements 
for brown dwarfs or ``super planets" mostly because
of the paucity of such objects, so we had to include ground-based
results. Along with the 
Kepler discoveries (Diaz et al. 2013; Diaz et al. 2014;
Bouchy et al. 2011; Moutou et al. 2013) we therefore included ground-based  results
(Johns-Krull et al. 2008; Hellier et al. 2009; Joshi et al. 2009; Siverd et al. 2012;
Anderson et al. 2011; Triaud et al. 2013).

Stellar masses and radii for most main sequence stars were taken from 
Torres, Andersen, \& Gim\'enez (2010). A variety of sources
were used for the parameters of stars from the low-mass end of the main sequence
(Pont et al. 2005, 2006, 2008; Demory et al. 2009; Tal-Or et al. 2013, Ofir et al. 2012;
Zhou et al. 2014).

Figure 1 shows the resulting mass-density relationship for our sample. There are two
major inflections in the curve. The maximum density occurs
at a mass of $\approx$ 60--70 $M_{Jupiter}$, 
the boundary between core nuclear burning stars
and degenerate core brown dwarfs. The second inflection is a minimum in density
at $M$ $\approx$ 0.3 $M_{Jupiter}$, roughly the boundary between H/He dominated planets
and low mass planets.

Fits were made to the data in these regions in order to define better
the boundary between lower mass planets, giant planets/brown dwarfs and stars. Although
the stars show a roughly linear relationship, in the
mass range 0.08 -- 1 $M_\odot$ the relationship is best fit by 
a second order polynomial  shown by the curved line.
For low mass stars in the range 
0.08 -- 1 $M_\odot$ the log $M$ $-$ log $\rho$ relationship can also be fit by a line
resulting in $\rho$ $\propto$ $M^{-(2.09 \pm 0.16})$. This follows from
the mass-radius relationship on the low mass end of the main sequence where 
$R$ $\propto$ $M$, thus $\rho$ $\propto$ $M^{-2}$.

The log $M$ -- log $\rho$ relationship for giant planets and brown dwarfs show a very
tight correlation (correlation coefficient, $r$ = 0.976). A linear fit over the mass
range 0.35  -- 65 $M_{Jup}$ results in:

$$log \rho  = (1.15 \pm 0.03) log M - (0.11 \pm 0.03) $$

This linear relationship of density with mass 
simply reflects the fact that objects with masses ranging from 
giant planets ($\sim$ 1 $M_{Jup}$) up to low mass stars all have approximately
the same radius. Thus an increase in 
mass is accompanied by proportional increase in density.
Note that this curve closely follows the mass-density relationship for H/He dominated
giant planets (Fortney et al. 2007) which is shown as the dashed line. 
At the low mass end of the giant planet range, 
the linear fit to the giant planets deviates significantly from the dashed line for planets with Jupiter-like composition. 
The planets below the dashed line are larger than expected by such models and called ``inflated" planets.
The linear relationship intercepts the main sequence for stars at $M$ = 63 $\pm$ 6 $M_{Jup}$.

For simplicity  we shall refer to all exoplanets with masses less than
$\approx$ 0.3 $M_{Jup}$ as low mass  exoplanets. These can be either 
rocky or ones that have a large fraction of volatiles (i.e. Neptune-like).  The low mass planets show considerable scatter
in their densities. In spite of this scatter there clearly appears to 
be a minimum in density around
$\sim$ 0.3 $M_{Jup}$. 
Indeed, a parabolic fit to the valley of this `V'-shape results in a minimum of the density at this value for the mass.
For higher mass objects the density increases.
We take this as the boundary between the low mass planets and 
giant planets.

\begin{figure}
\plotone{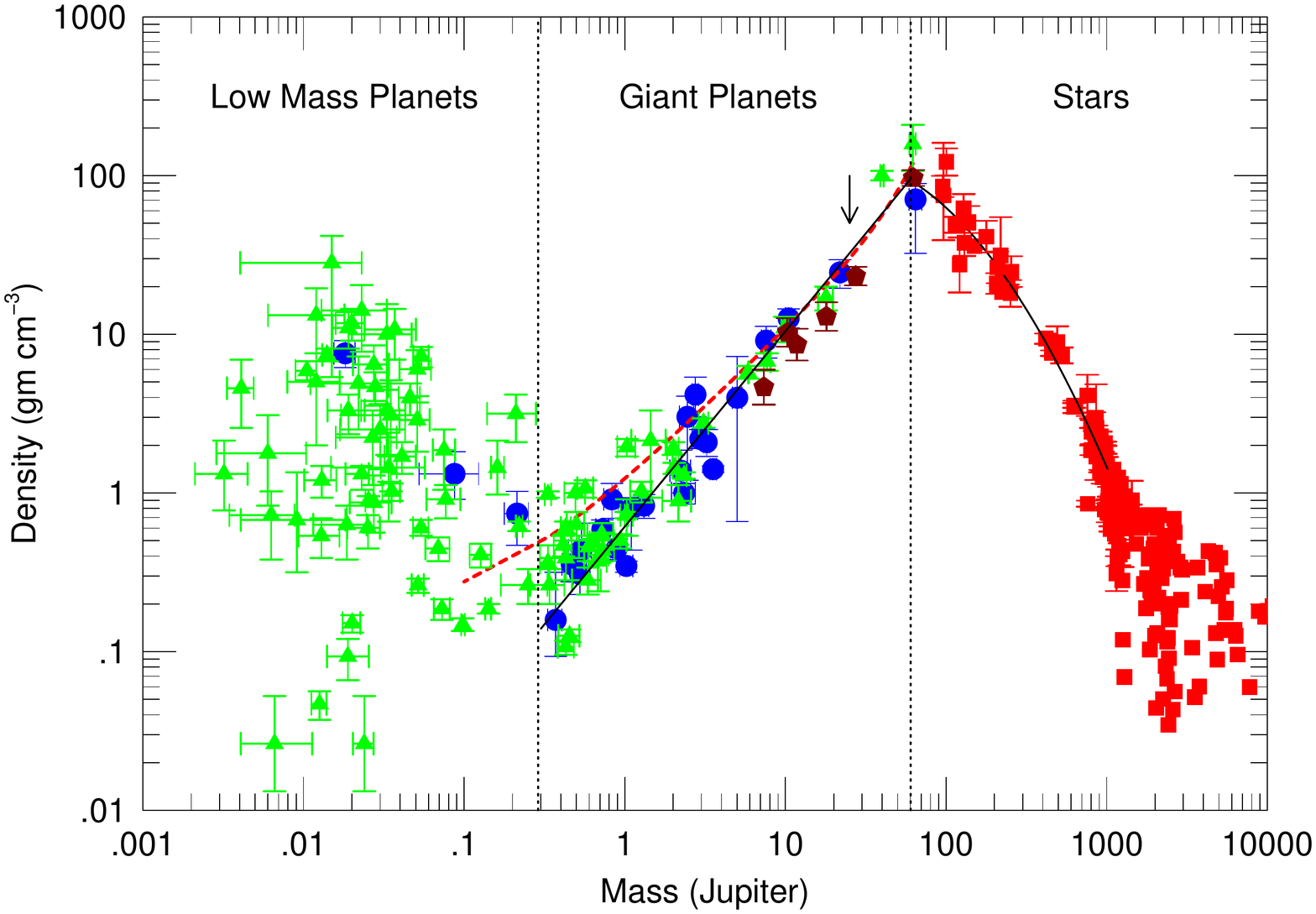}
\figcaption[ftall.ps]{The density and mass of stars (red squares), giant planets and brown
dwarfs, and low mass planets. Triangles represent Kepler discoveries and dots are CoRoT exoplanets. Ground-based
discoveries for high mass giant planets are shown by pentagons.  The line 
represents a linear fit to the giant planets and brown dwarfs
in the mass range $M$ = 0.35 - 60 $M_{Jup}$. A second order polynomial fit (curved line) was made
to the lower end of the stellar main sequence. The boundary between the low mass planets
and giant planet occurs at $M$ $=$ 0.3  $M_{Jup}$. The boundary between the giant planets
and stars is at $M$ $=$ 60 $M_{Jup}$ (0.060 $M_\odot$). The dashed red line shows the mass-density
relationship for H/He dominated giant planets taken from Fortney et al. (2007).
\label{f1}}
\end{figure}

\begin{figure}
\plotone{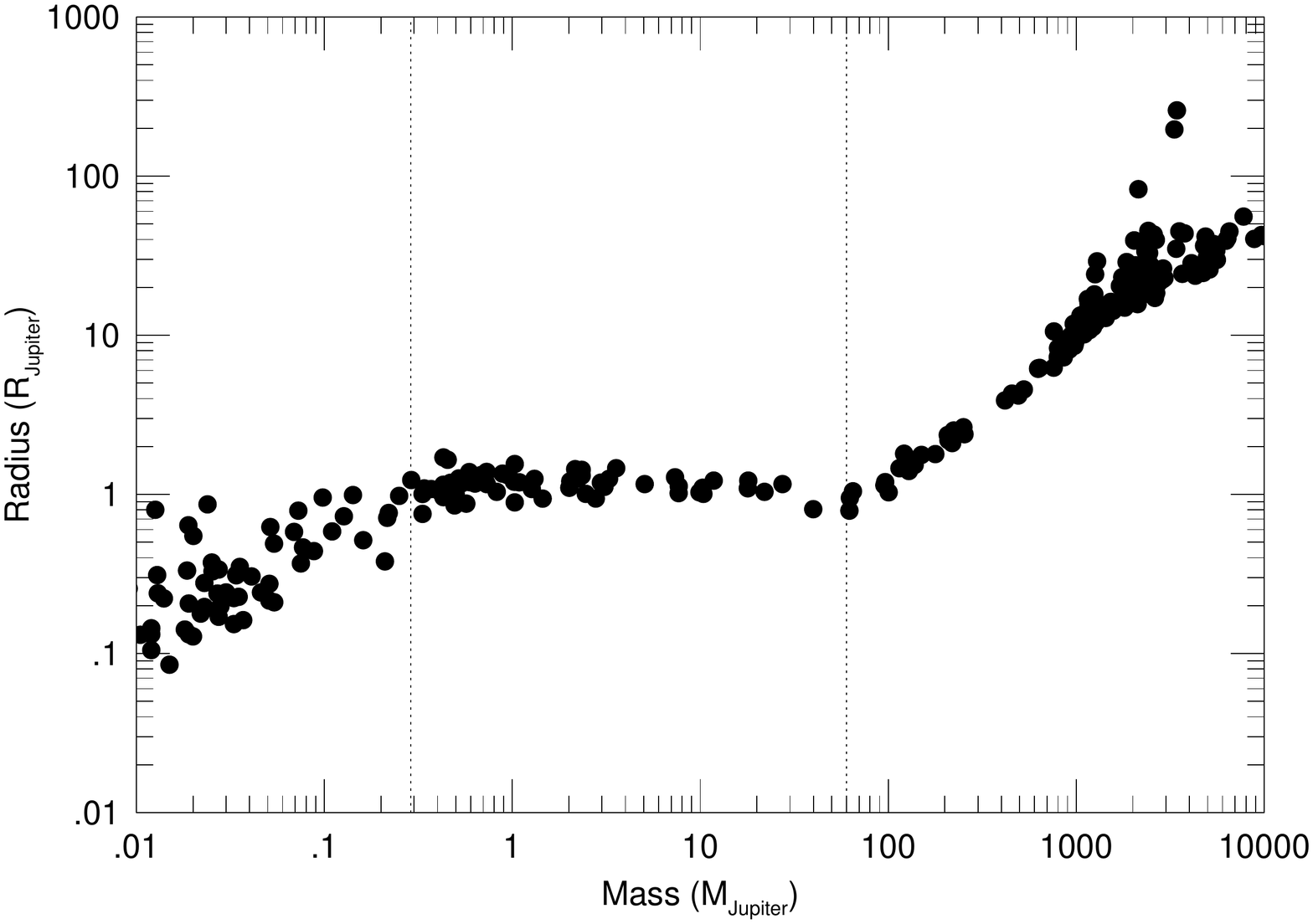}
\figcaption[ftall.ps]{The points from Figure 1 shown in the mass-radius plane.
\label{f2}}
\end{figure}

Fig.~\ref{f2} shows the more traditional Mass-Radius relationship with our boundaries 
shown as the vertical dashed lines. 
One can also see
inflections in this curve at roughly the same boundaries seen in Fig.~\ref{f1}, although these are not as striking or well 
defined particularly between the low mass and giant planets.

\section{Discussion}

The mass density diagram for all objects with masses
ranging from planets to stars are 
separated by three distinct regions marked by an abrupt change in sign of the
slope of the $\log$\,M $-$ $\log \rho$  relationship. Stars
show a negative slope, whereas giant planets and brown dwarfs  have a positive slope.
However, below about  0.3  $M_{Jup}$ objects show considerable scatter in their
densities. We shall simply refer to this region of the diagram as the ``low mass planets" (LMP).
It is beyond the scope of this paper to discuss this region of the diagram, the origin of
such scatter, or any relationship between the mass and density. Instead we will focus
on the giant planets and 
for  the sake of discussion we shall refer to these regions between the LMP and stars as 
 the ``gaseous planet sequence'' (GPS).  MS will refer to the classic  
main sequence for stars. 

The beginning of the  GPS
occurs at $M$ $\approx$  0.3  $M_{Jup}$, roughly the boundary between planets with a significant
amount of volatiles, and those dominated by H/He.   The boundary  between the
GPS and the MS occurs at 60 $M_{Jup}$.   
Laughlin \& Lissauer (2015) noted that the distribution of planetary densities
had a broad minimum at a planet mass of $M_P$ $\sim$ 0.1 $M_J$
(30 $M_{Earth}$) which they took as the boundary between
giant planets and low mass planets (termed ``ungiants'' in their paper)
We propose that this boundary
is actually at a much higher mass of  $M_P$ $\sim$ 0.3 $M_J$.

We also fit the density-mass data from Laughlin \& Lissauer (2015, Fig. 5 of their paper)
for planets with in the mass range 0.01 to  0.1 $M_J$. 
The data are well fit by a linear
function that intercepts our GPS at $M_P$ $\sim$ 0.3 $M_J$ ($\approx$ 100 $M_{Earth}$), 
consistent with our proposed
boundary.

 The striking
feature about GPS  is that there is  no distinguishing characteristic
that separates the 
low mass end where objects are clearly planets, and the high mass end 
where objects are generally considered to be brown dwarfs. The 25 $M_{Jup}$ limit
(the arrow in Fig. 1) taken by Schneider et al. (2009) to be the  boundary between
planets and brown dwarfs shows no obvious differences in the GPS 
on either side of this limit.  If anything, the arrow only seems to mark the boundary where
the data are sparce. 
Clearly, the discovery of more objects
in this mass range is desperately needed. Possibly differences between the
giant planets and the ``traditional'' brown dwarfs may  become more apparent
with more discoveries. For instance, if no objects can be found
that fill the gap marked by 25 $M_{Jup}$ (arrow) and the onset
of the MS then this ``gap'' might be taken to separate the planets from the
brown dwarfs and stars. For now we note that the few brown dwarfs with
masses $\approx$ 60 $M_{Jup}$ all fall on the GPS.

In light of Figure 1 making a distinction  between objects on the low-
and high-mass end of the GPS
seems arbitrary and may only obscure the fact 
that all objects on this track are physically the same objects governed by the same
underlying physics and with similar structure, analogous to the MS.
Chabrier et al. (2014) also argued that the mass boundary 
between giant planets and brown dwarfs given by the present IAU 
definition was ``incorrect and confusing and should be 
abandoned''. Figure~\ref{f1} certainly supports this claim. This figure
also shows that the density provides us
with no obvious hints regarding a different formation mechanism between brown dwarfs
and giant planets.  Possibly the differences in the physical properties of the atmospheres
may show indications of a different formation mechanism (Chabrier et al. 2014).

Comparing the properties of the GPS to the MS provides us with additional arguments
to support the claim that all objects along the GPS should
be considered the same type of objects. Stars have masses that cover over
two orders of magnitude. The structure of the star changes considerably
along the main sequence. High mass stars have a convective core and radiative envelope and
as one moves down the main sequence this structure changes to objects with a radiative core
and a convective envelope. The lowest mass stars, on the other hand, are fully 
convective. The stellar atmosphere changes considerably along the main sequence
in terms of effective temperatures and the types of spectral features that are
observed. 

Finally, stars may also have different formation 
scenarios. Low mass stars are believed to form from the collapse of a proto-cloud and 
subsequent accretion (Palla \& Stahler 1993). The
formation mechanism for high mass stars is still open to debate. For massive stars
radiation from the core halts the accretion process thus limiting the mass (e.g. Yorke \&
Kr{\"u}gel 1977). 
One hypothesis is that they are formed by the merger of lower mass stars (Bonnell et al. 1998).
Regardless of all these substantial differences, all objects along the
MS are considered  to be the same general class of objects that is governed
by the same physics - nuclear burning in the core under hydrostatic equilibrium. We only
make sub-distinctions in the form of ``low mass'' and ``high mass'' stars.

Objects along the GPS also have masses that differ by over two
orders of magnitudes. Like stars, they certainly have a wide range
of effective temperatures, atmospheric features, and possibly even
different
formation mechanisms. Making an arbitrary distinction between giant planets
and brown dwarfs only confuses the central issue that these objects share
a similar structure. 
Analogous to stars we can make
subtle distinctions between ``high mass giant planets'' and ``low mass giant
plants'', but the class should be considered as a whole in order to gain
a more fundamental understanding of their formation, evolution, and nature.

Comparing the GPS to the MS also provides a natural explanation to the
so-called ``Brown Dwarf Desert". The paucity of brown dwarf companions
simply reflect the decrease in number of high mass 
much in the same 
way there is a decrease in the number of high mass stars in the stellar
distribution.
Compared to low mass main sequence stars
O-type stars in our galaxy are extremely rare -- it is simply
harder for nature to form these higher mass stars. To our knowledge, astronomers never
refer to the ``O-star desert."

We propose a new definition of planets, brown dwarfs, and stars based not on 
arbitrary separation of distributions, or whether short-lived deuterium
burning has occurred, or just because we are biased in thinking that giant planets
should all have  masses close to that of our Jupiter. Rather, our definition is  based
on the observed inflections in the mass-density diagram that separate regions governed
by different underlying physics. Thus,

$M$ $<$ 0.3  $M_{Jup}$  $\Rightarrow$ Low Mass Planets

0.3  $M_{Jup}$ $<$ $M$ $<$ 60 $M_{Jup}$  $\Rightarrow$ Giant Gaseous Planets

$M$ $>$ 60  $M_{Jup}$  $\Rightarrow$ Stellar Objects

We note that by our definition Saturn has a mass near
the boundary between low mass planets and gas giant planets.
Although we refer to objects with $M$ $>$ 60  $M_{Jup}$   as 
``stars'', the exact boundary between objects supported by electron degeneracy pressure
and those  with a hydrogen burning core is not
well known and can be as high as 80 $M_{Jup}$. Possibly objects with 
60 $M_{Jup}$ $<$ $M$ $<$ 80 $M_{Jup}$ should be considered to be the
bona fide brown dwarfs.

Another obvious feature  of Figure 1 is the relative paucity of objects in
the mass range 20 $<$ $M_{Jup}$ $<$ 100. This is in part due to the relative low
number of these with respect to lower mass planets, but may also be due to the fact
that Doppler surveys have largely concentrated on confirming transit discoveries
of lower mass planets. Low mass stars, and objects on the high mass
end of the GPS are often ignored in favor of getting mass measurements
on the more ``interesting'' planet candidates. However, accurate mass and radius
measurements of objects on the low mass end of the stellar main sequence
and the upper end of the GPS, i.e. the boundary between high mass giant
planets and low mass stars, are also important. Only by studying the full
range of objects from high mass giant planets to the lower end of the main
sequence will we obtain a more fundamental understanding of the formation
of giant planets compared to low mass stars.

In much the way the Hertzsprung-Russel Diagram has served as a tool
for understanding stellar structure, the M-$\rho$ diagram
can serve as a powerful tool for understanding planetary structure.

\end{document}